\documentclass[11pt,a4paper]{article}
\usepackage[top=0.5in,bottom=1.0in,left=1.0in,right=1.0in]{geometry}

\usepackage{multirow}
\usepackage{multicol}
\usepackage{amsthm}
\usepackage{makecell}
\usepackage{supertabular}
\usepackage{longtable}
\usepackage{siunitx}
\usepackage{authblk}
\usepackage{numprint}
\usepackage{graphicx}   
\usepackage{float}
\usepackage{xurl}

\usepackage{csquotes}
\usepackage[natbib=true,
            style=ieee,
            sorting=none,
            citestyle=numeric-comp
            ]{biblatex}
\title{Exploration of Characteristic Temperature Contributions to Metallic Glass Forming Ability}

\author[1]{Lane E. Schultz}
\author[1]{Benjamin Afflerbach}
\author[1]{Carter Francis}
\author[1]{Paul M. Voyles}
\author[1]{Izabela Szlufarska}
\author[1]{Dane Morgan}

\affil[1]{University of Wisconsin-Madison, 1500 engineering Drive, Madison, WI 53706, USA}

\begin{document}
\maketitle







\begin{abstract}
Various combinations of characteristic temperatures, such as the glass transition temperature, liquidus temperature, and crystallization temperature, have been proposed as predictions of the glass forming ability of metal alloys. We have used statistical approaches from machine learning to systematically explore a wide range of possible characteristic temperature functions for predicting glass forming ability in the form of critical casting diameter, $D_{max}$. Both linear and non-linear models were used to learn on the largest database of $D_{max}$ values to date consisting of 747 compositions. We find that no combination of temperatures for features offers a better prediction of $D_{max}$ in a machine learning model than the temperatures themselves, and that regression models suffer from poor performance on standard machine learning metrics like root mean square error (minimum value of $3.3 \pm 0.1$ $mm$ for data with a standard deviation of 4.8 $mm$). Examination of the errors vs. database size suggest that a larger database may improve results, although a database significantly larger than that used here would likely be required. Shifting a focus from regression to categorization models learning from characteristic temperatures can be used to weakly distinguish glasses likely to be above vs. below our database's median $D_{max}$ value of 4.0 $mm$, with a mean F1 score of $0.77 \pm 0.02$ for this categorization. The overall weak results on predicting $D_{max}$ suggests that critical cooling rate might be a better target for machine learning model prediction.
\end{abstract}





\section{Introduction}

\par
Physically motivated models built using powers and ratios of sums and differences of experimental measures of the glass transition $T_{g}$, onset to crystallization, $T_{x}$, and liquidus, $T_{l}$, temperatures (so-called characteristic temperatures (CTs)) have been explored for decades to predict metallic glass forming ability (GFA). We will call these powers and ratios of sums and differences PRSD functions in this paper. There exist several quantitative measures of GFA. The critical cooling rate, $R_{c}$, is the slowest a molten metal can be cooled to produce a glass. The smallest dimension of the largest glassy sample for a composition is defined to be $Z_{max}$ whereas the maximum rod diameter of a glassy specimen manufactured through suction casting is the critical casting diameter, $D_{max}$. Both $D_{max}$ and $Z_{max}$ denote the maximum reachable thickness for a glass but differ by geometry. Hence, predictions on either $D_{max}$ or $Z_{max}$ denote the ability for a model to quantify the maximum thickness for a glassy metal sample. Starting in 1969, the ratio between $T_{g}$ and $T_{l}$ was introduced to quantify the ease of forming a bulk metallic glass \cite{Turnbull1969}. This was the original form of the reduced glass transition temperature, $T_{rg}=T_{g}/T_{l}$. Attempts to use the melting temperature instead of $T_{l}$ were made but resulted in worse models for $log_{10}(R_{c})$ \cite{Lu2000}.

\par
More modern efforts to model GFA with CTs have only modestly increased in complexity and still generally focus on correlating a metric of GFA with PRSD functions of CTs. For instance, $\gamma=T_{x}/(T_{g}+T_{l})$ was introduced in 2002 and is a single term ratio between two quantities \cite{Lu2002}. The $\gamma$ parameter was constructed by using the devitrification range, $(T_{x}-T_{g})$, and simplifications based on classical crystal growth and nucleation theory. Lu and Liu \cite{Lu2002} showed that $\gamma$ shows a strong relationship with $log_{10}(R_{c})$ ($R^{2}=0.91$) but much weaker correlation with critical section thickness $Z_{max}$, ($R^{2}=0.57$).

\par
In 2005, another parameter based on PRSD functions of CTs was introduced as $\alpha=T_{x}/T_{l}$ \cite{Mondal2005}. The difference between this parameter and $T_{rg}$ was the substitution of $T_{x}$ for $T_{g}$. In the same work, $\beta=T_{rg}+T_{x}/T_{g}$ was also introduced. The $R^{2}$ scores for $\alpha$ and $\beta$ were 0.90 and 0.93 against $log_{10}(R_{c})$ respectively. Similar to the parameter $\gamma$, both $\alpha$ and $\beta$ degraded in $R^{2}$ performance for prediction on $Z_{max}$, with $R^{2}$ of 0.48 and 0.54, respectively.

\par
Another PRSD function of CTs used as a feature for GFA is $\omega=T_{g}/T_{x}-2T_{g}/(T_{g}+T_{l})$ introduced in 2008 \cite{Long2009}. This parameter takes into account the devitrification range and liquid stability. With an $R^{2}$ of 0.93 against $log_{10}(R_{c})$ for 53 metallic alloys, $\omega$ provides the best performing model for $log_{10}(R_{c})$ of which we are aware to date. Since $D_{max}$ should increase with a decreasing $R_{c}$, the comparison between $1/\omega$ and $D_{max}$ was performed in the same work. The $R^{2}$ was 0.41 which follows previous performance trends of relatively poor $R^{2}$ of PRSD functions of CTs for $D_{max}$ even when good performance for $R_{c}$ was obtained.

\par
A common theme throughout attempts to model GFA from CTs is that they are based on linear models of PRSD functions, and usually just a single PRSD function. They also have a tendency to do well with learning on $log_{10}(R_{c})$ but not on $D_{max}$ or $Z_{max}$. See Refs.~\cite{Deng2020, Xiong2019} for more comparisons between these types of linear models and their correlations against $D_{max}$.

\par
Recently, there have been machine learning (ML) efforts to learn $D_{max}$ from CTs that go beyond simple linear correlations with PRSD functions. Because of the availability of more $D_{max}$ data compared to $R_{c}$ measurements, more advanced ML techniques can be implemented and reliably assessed for learning on $D_{max}$. Specifically, Xiong, et al. \cite{Xiong2019} used a Gaussian process (GP) model to learn $D_{max}$ from CTs. The study predicted $D_{max}$ on 442 metallic glasses with an $R^{2}$ of 0.76 for the training set. Even more recently, Deng and Zhang \cite{Deng2020_2} used the random forest (RF) method to learn $D_{max}$ from CTs and some other features for the same dataset as used by Xiong, et al. \cite{Xiong2019}. Deng and Zhang found an $R^{2}$ of 0.64.

\par
Thus far, the largest ML attempt to quantify $D_{max}$ as a function of CTs was done by Xiong, et al. \cite{Xiong2020} with 674 compositions. RF models were trained on the three CTs mentioned in this study. They used 100-fold cross-validation (CV) where an RF model was trained on 99 folds while its performance was measured by the leave out set.  Their model had an $R^{2}$ of 0.60 ($R=0.77$) and a root mean squared error ($RMSE$) of 2.89 $mm$. However, this result was attained by excluding six erroneous compositions which showed large residuals in $D_{max}$ predictions for another one of their models. In the spirit of generating a PRSD function of CTs, Xiong, et al. also used symbolic regression with CTs to generate a three-term model on their dataset (Equation~\ref{symbolic_function}). The symbolic regression model scored an $R^{2}$ of 0.45 ($R=0.67$) and an $RMSE$ of 3.37 $mm$ \cite{Xiong2020}. Whether PRSD functions of CTs can aid in quantifying $D_{max}$ through other ML methods remains an open question.

\begin{equation}\label{symbolic_function}
    D_{max} = f \Biggl( \frac{T_{g}^{2}}{T_{l}^{3}}, \\
              \frac{1940}{136263T_{l}-1426T_{l}(T_{x}-T_{g})}, \\
              \frac{T_{g}^{2}}{39T_{l}(14T_{l}-(T_{x}-T_{g})^{2})} \Biggl)
\end{equation}

\par
While the ML approaches to learning $D_{max}$ from CTs appear to be performing better than linear PRSD functions of CTs, many ML studies to date suffer from several shortcomings. First, model performances tend to be measured with $R^{2}$. Other error metrics such as mean average error ($MAE$) and $RMSE$ are generally not provided. Second, and more importantly, the reported $R^{2}$ for some previous efforts are for predicting back on training data, rather than assessment on test data not seen in the fitting process. Thus, there was no assessment of the extent to which trained models have predictive power outside their training sets. Having just training data results is common when modeling $D_{max}$ using simple linear models of PRSD functions of CTs, and while this may lead to some overfitting and underestimation of errors, the simplicity of these models may make these underestimations negligible. However, more complex ML are particularly subject to overfitting and careful assessment with test data is essential for robust assessment of predictive ability.

\par
Our work focuses on providing multiple metrics of assessment for ability to predict on $D_{max}$ on carefully excluded validation data, avoiding data leakage through nested cross-validation. Our approach generalizes previous use of select PRSD functions as features by generating a comprehensive set of PRSD functions up to reasonable powers. We then explore the efficacy of these features in multiple ML model types, including least absolute shrinkage and selection operator (LASSO), Gaussian kernel ridge regression (GKRR), RF, and Gradient Boosting (GB). The benefit of using ensemble models is the inherent feature selection that they provide. The L1 norm for LASSO tends to penalize arbitrary feature weights to zero which is also a form of feature selection. This approach can therefore assess whether features based on PRSD functions of CTs yield effective predictive models of $D_{max}$ and to what extent the use of PRSD functions provide better predictions than just using the CTs themselves. We also test whether learning on the $log_{10}(D_{max})$ is more effective than learning on $D_{max}$ and if applying principal component analysis (PCA) to transform our features had any added benefit. Our dataset is comprised of 747 compositions, which is the largest set to date used for building and assessing models to predict $D_{max}$.

\section{Materials and Methods}\label{Methods}

\par
Experimental data was provided by Ref.~\cite{metallic_glasses_data}. Measurements of $D_{max}$ are susceptible to differences in experimental setup which could impact cooling rate.  The experiments in the database were all melt quenched using similar rod like molds and melting processes that are standard among experimentalists. Only integer $D_{max}$ values are reported which introduces small uncertainties. The characteristic temperature $T_{l}$ should be unaffected by heating and cooling rates but $T_g$ and $T_x$ are heating and cooling rate dependent.  In the database, entries were not constrained to certain heating and cooling rates, but the majority of the heating rates fall into the range of 10-100 degrees per minute as this is fairly standard procedure for the differential scanning calorimetry (DSC) used to measure these values. The impact of variation in heating and cooling rates on the ML accuracy is difficult to asses and is an interesting topic for further study but we have not attempted to explore it here.

\par
A set of unique compositions provided by Ref.~\cite{metallic_glasses_data} with values of $T_{g}$, $T_{x}$, $T_{l}$, and $D_{max}$ were used to generate features. First, differences and summations between pairs of temperatures were taken (e.g., $(T_{g}+T_{x})$ and $(T_{l}-T_{g})$). These features were then raised to the powers of -4 to 4 (e.g., $(T_{g}+T_{x})^{2}$ and $(T_{l}-T_{g})^{-3}$). Products between all aforementioned features were then included to produce the feature set of PRSD functions of CTs (e.g. $(T_{g}+T_{x})^{2} \cdot (T_{l}-T_{g})^{-3}$). Any instance that resulted in a division by zero was eliminated from the analysis. For compositions that appeared more than once in the database and had a full set of the three CT values, the maximum value for $D_{max}$ and the mean values for CTs were used. However, if any one instance of a CT was more than 50 $K$ from the mean, then that value was excluded and the mean was taken with the remaining points to minimize erroneous CTs values. This data processing reduced the complete database of 6,914 entries to 747 unique compositions, each with a $D_{max}$ value, three unique characteristic temperatures ($T_{g}$, $T_{x}$, $T_{l}$) and 2,628 features from PRSD functions of CTs described above. The processed data can be found in Ref.~\cite{GB_model} and Ref.~\cite{Morgan2021}.

\par
Features were standardized to have a zero mean and unit variance. A separate feature set was generated by applying principal component analysis (PCA) while keeping all principal components to transform features. Some of the PRSD functions of CTs features are highly correlated which means that they provide similar information. PCA can be used to transform a feature set to a lower dimension or to an equivalent set with linearly independent features. The principal  components are orthogonal vectors and explain the maximum variance of data along several directions. The larger a singular value for a corresponding principal  component, the more data that principal  component represents from the original dataset. Therefore, data sets with linearly dependent features can be transformed to a linearly independent data set with PCA if all nonzero principal components are kept for the projection.

\par
In ML, having more features (explanatory variables) than observations when building a model can lead to overfitting. An overfit model generalizes poorly and is less likely to correctly predict a target variable for cases withheld from training. The degree of overfitting tends to increase with the number of features included in training. Because the number of generated features in the present study are much larger than the number of observations, we apply models that reduce the number of contributing features via regularization (LASSO), shrinkage (GB), and bagging (RF). We will show that a subset of features that number less than the number of observations contribute to predictions by a GB model (Section~\ref{res_desc}) which is likely to hold true for LASSO and RF model types as well.

\par
All assessment done used a nested CV approach \cite{Cawley2010}. The nested CV used a 5-fold inner and 5-fold outer loop. A grid search of hyperparameters was applied for the inner loop in the nested CV (Table~\ref{gridsearch}) to establish optimal hyperparameters for each outer loop fold. Fitting was then done for each outer CV fold on the full training set not in the test fold ($80\%$ of the data). Then, the model was applied to the outer CV test set fold. To test whether the generated features provide improved prediction, we also applied nested CV with only $T_{g}$, $T_{x}$, and $T_{l}$. Data were randomized every time nested CV was applied, meaning that the splits for the testing and training sets may have differed for comparisons. All ML was performed with scikit-learn \cite{scikit-learn}.

\begin{table}[H]
\centering
\caption{The hyperparameter grid for each model type explored is tabulated below. Variable names follow the convention of scikit-learn \cite{scikit-learn}.}
\begin{tabular}{|c|c|c|}
\hline
{Model}                 & {Parameter}   & {Values}        \\ \Xhline{3\arrayrulewidth}
LASSO                 & alpha                      & \makecell{100 values \\ from 0 to 5 in log10 space}  \\ \hline
\multirow{3}{*}{GKRR} & alpha                      & \makecell{100 values \\ from -5 to 5 in log10 space} \\ \cline{2-3} 
                               & kernel                     & rbf                                    \\ \cline{2-3} 
                               & gamma                      & \makecell{100 values \\ from -3 to 3 in log10 space} \\ \hline
\multirow{3}{*}{RF}   & n\_estimators       & 30, 40, 50, 60, 100, 500               \\ \cline{2-3} 
                               & max\_features & sqrt, log2, None                       \\ \cline{2-3} 
                               & max\_depth              & 2, 3, 4, None                          \\ \hline
\multirow{4}{*}{GB}   & learning\_rate              & 0.001, 0.01, 0.1, 0.2                  \\ \cline{2-3} 
                               & n\_estimators       & 30, 40, 50, 60, 100, 500               \\ \cline{2-3} 
                               & max\_features & sqrt, log2, None                       \\ \cline{2-3} 
                               & max\_depth              & 2, 3, 4                                \\ \hline
\end{tabular}
\label{gridsearch}
\end{table}

\par
For LASSO, GKRR, RF, and GB, the following set of four tests were performed: fit with non-PCA features and $D_{max}$, fit with PCA features and $D_{max}$, fit with non-PCA features and ~$log_{10}(D_{max})$, and fit with PCA features and $log_{10}(D_{max})$. $RMSE$, $MAE$, $R^{2}$, and $RMSE/\sigma$ were calculated for each fold in the outer loop from the nested CV (where $\sigma$ is the standard deviation of the true target values in that fold). This gave five values for each metric for each nested CV run, and we performed one nested CV run for each test, for a total of 5 values for each metric (i.e., 5 values of $RMSE$, 5 values of $MAE$, etc.). These distributions of values for each metric were used to find the mean value, standard deviation (STDEV), and standard error in the mean (SEM), for each metric from each nested CV. All metrics with units are in units of $D_{max}$, which is in millimeters ($mm$). If fitting was done with $log_{10}(D_{max})$, then the predicted output was transformed back to $D_{max}$ before calculating error metrics. All metrics are scores from outer folds of nested CV runs unless explicitly stated otherwise.

\par
Separate from assessing the accuracy of our models, we use nested CV, we would also like to develop the most complete and accurate model possible using the whole database. To do this a GB model was trained using all the data. The optimal hyperparameters were found by applying a grid search using 5-fold CV on the whole data set and call this model GB~1. From GB~1, we can assess which of the generated features provide the most utility for regression prediction. We studied the impact of fitting GB models with the top $n$ features by building a learning curve with 5-fold CV and measuring $RMSE/\sigma$. The curve was averaged over the leave out sets. The uncertainties are in SEM. The choice of hyperparameters were kept from GB~1. The 50 highest ranking features were used to fit a final GB model, named GB~2, because regression performance did not significantly change by including the remaining features. GB~2 can be found at the Materials Data Facility (MDF) online data and code sharing repository at Ref.~\cite{GB_model} and figshare at Ref.~\cite{Morgan2021}.

\par
To test whether the generated features had a significant impact on learning, we performed a two-sided T-test for the distribution of five scores for each metric obtained above using all model types. One distribution of scores were from the generated features while the others were from using just the three CTs as features. Both feature sets did not have a PCA transformation and learned from $D_{max}$ directly. The aforementioned comparison choice  was performed because models tended to degrade in performance with PCA transformed features and with application of a logarithm onto $D_{max}$.

\par
For LASSO models using PRSD functions of CTs, models had unusually poor regression performance on some test folds for nested CV. Many generated LASSO models were not well conditioned due to numerical problems and gave outlandish predictions. As a result, an outer fold was removed if any of the regression metric values were outside of a multiple of 3 from the optimal GB workflow metrics. This left three outer folds for any LASSO metric reported for all combinations of logarithm application on $D_{max}$ and PCA application on the feature set. Due to the low number of individual observations, LASSO models were excluded from the two-sided T-tests. All LASSO models trained during hyperparameters grid searches showed acceptable convergence.

\par
A learning curve was generated to test if improved learning would result from more data. Nested CV was performed for the best GB workflow for $10\%$ up to $100\%$ for the 747 compositions by increments of $10\%$. Each subset of data was randomly sampled. Learning curves were built from the predictive performance of the testing sets from nested CV.

\par
To test if the generated features had predictive power for classification, nested CV was performed using a GB classifier. No PCA was applied nor a logarithm onto $D_{max}$. Any composition with a $D_{max}$ less than $4.0$ $mm$ were assigned to be class 0 while all others were assigned class 1. The median $D_{max}$ value for the dataset was $4.0$ $mm$ which splits the classes evenly. Binary classification scores for every test set in nested CV were gathered.

\par
We mark our contribution to the field by comparing our best performing regression workflow to the work done by Xiong, et al. in Ref. \cite{Xiong2020}. We make sure to match the number of folds used in their CV and the same set of features. Additionally, we attempt to predict $D_{max}$ with our dataset using their reported model shown in Equation~\ref{symbolic_function}. We use ordinary least squares with three fitting coefficients and one intercept term to fit Equation~\ref{symbolic_function} to all of our data. The least squares model was then used to predict back on our full dataset.

\section{Results and Discussion}\label{res_desc}

\par
The lowest $RMSE/\sigma$ model types were GB and RF from learning on the generated feature set without PCA and using $D_{max}$ (rather than $log_{10}(D_{max})$) as the target feature. The mean of $MAE$ was slightly lower for the optimal GB workflow (Tables~\ref{gen_mean_metrics_gb}-\ref{gen_mean_metrics_rf}). The lowest mean ~$RMSE/\sigma$ ($MAE$) for GB and RF were 0.70 (2.18 $mm$) and 0.70 (2.24 $mm$) respectively. For all model types, it was found that fitting on $D_{max}$ instead of $log_{10}(D_{max})$ gave better performance for mean $RMSE/\sigma$ from nested CV. Application of PCA was better for GKRR and LASSO, and even in that case the effect was small. All values are tabulated on Table~\ref{gen_mean_scores}. Scores for using the just the three CTs as features are provided in Tables~\ref{raw_mean_metrics_gb}-\ref{raw_mean_metrics_rf}.

\begin{table}[H]
\centering
\caption{The mean $RMSE/\sigma$ scores for each test for each model type along with their standard deviations (STDEV) and standard error in the mean (SEM).}
\npdecimalsign{.}
\nprounddigits{2}
\begin{tabular}{|c|c|c|n{5}{2}|n{5}{2}|n{5}{2}|}
\hline
{Model} & {$Log_{10}$} & {PCA} & {Mean} & {STDEV} & {SEM} \\
\hline
GB    & False & False & 0.6999509862       & 0.0801319762526573   & 0.035836109214468115 \\
GB    & False & True  & 0.857365730626731  & 0.10079693444412109  & 0.04507775946812895  \\
GB    & True  & False & 1.3682035770070058 & 0.037107466404276084 & 0.016594963470550204 \\
GB    & True  & True  & 1.3800803923241676 & 0.061462271163876564 & 0.02748676327479062  \\
GKRR  & False & False & 0.7553606795917878 & 0.06282084870501099  & 0.02809433762172684  \\
GKRR  & False & True  & 0.7522565114170061 & 0.0865628482612408   & 0.038712082607626784 \\
GKRR  & True  & False & 1.3854035599885715 & 0.0886005402698009   & 0.03962336617729647  \\
GKRR  & True  & True  & 1.3772221668571007 & 0.05334841742827817  & 0.0238581375723329   \\
LASSO & False & False & 0.8648449751236899 & 0.028263517712638183 & 0.01631794955963741  \\
LASSO & False & True  & 0.8148350498666646 & 0.04149065128203427  & 0.023954638686535377 \\
LASSO & True  & False & 1.3995895937524139 & 0.04603859480975342  & 0.026580395106523245 \\
LASSO & True  & True  & 1.3602218043301997 & 0.03180678290260179  & 0.01836365467087313  \\
RF    & False & False & 0.6979387534177431 & 0.04267062132404659  & 0.01908288198454405  \\
RF    & False & True  & 0.9203845318995703 & 0.1434376447785077   & 0.06414726485144219  \\
RF    & True  & False & 1.372737087333297  & 0.059896939007054326 & 0.026786725452786445 \\
RF    & True  & True  & 1.3817012508845425 & 0.05315296734904437  & 0.023770729639658    \\
\hline
\end{tabular}
\label{gen_mean_scores}
\end{table}

\par
\begin{figure}[H]
\centering
\includegraphics[width=0.75\textwidth]{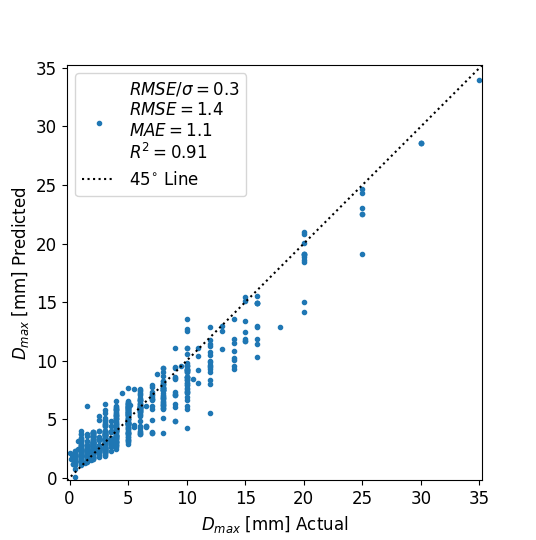}
\caption{The parity plot for training set prediction for GB~2 is shown.}
\label{parity_model}
\end{figure}

GB~2, which was trained on the top 50 ranking PRSD functions of CTs, is shown in Figure~\ref{parity_model}. GB~2 was attained by using the optimal hyperparameters from GB~1 and showed outstanding performance on regression metrics $RMSE/\sigma=0.30$ and $R^{2}=0.91$ (not from nested CV). While GB~2 is taken as the best overall model for predicting new data as it is fit and optimized on all our present data, the 5-fold CV performance of GB~2 cannot be taken as predictive for new data due to data leakage and overfitting. The error metrics from the nested CV are the best predictor of the expected performance on new data for GB model types.

\par
\begin{figure}[H]
\centering
\includegraphics[width=0.75\textwidth]{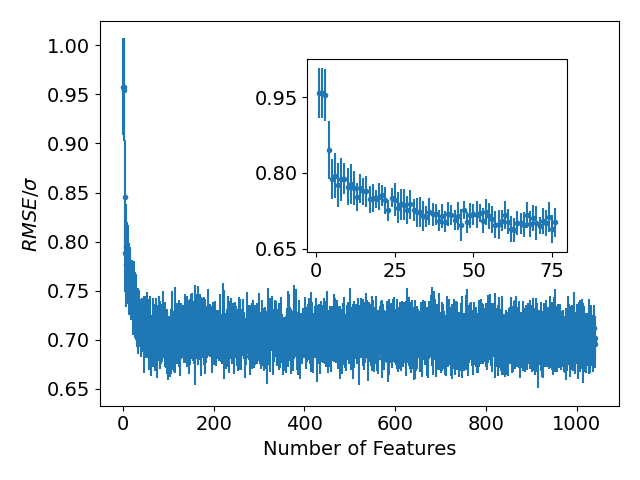}
\caption{The number of features included in GB models as a function of $RMSE/\sigma$.}
\label{gb_features_loss}
\end{figure}

GB~1 ranked 1,042 out of the 2,628 features to be of some nonzero significance. A subset of those features with scores are tabulated in the Appendix. This is a very large number of features, even more than the number of data points, and suggests that the model is very poorly constrained. However, the actual number of significant parameters that impact the model is likely far fewer. To asses the truly significant parameters, GB models were fit incrementally with subsets of features, starting with the highest ranking, then the top two ranking, then the top three ranking, and so on. The GB models used the GB~1 hyperparameters described in the Methods section. $RMSE/\sigma$ with respect to the number of included features, ordered by their ranking, is shown in Figure~\ref{gb_features_loss}. There is essentially no gain in in prediction performance after about 50 features. Thus, a GB model represented with all the features can be equally represented with just the first 50 features as these are all that really contribute to its accuracy. In general, it is a concern when one has more fitting parameters than data points. Although models mentioned in this work can formally provide fits when they have more fitting features than cases, only a number much less than 747 are found in this work to contribute to predictions when training on PRSD functions of CTs as shown by GB~2.

\par
The p-values for the two-sided T-test for comparing results from using features that are PRSD functions of CTs and just the three CTs are reported in Table~\ref{p-vals}. None of the values for all reported metrics fell below 0.32, far from the value of 0.05 typically used as a cutoff to claim significant difference. Hence, there was no statistically significant difference between learning from the generated features versus the original CTs.

\begin{table}[H]
\centering
\caption{The p-values for comparing models based on generated features and just three CTs for each ML scoring metric.}
\npdecimalsign{.}
\nprounddigits{2}
\begin{tabular}{|c|n{5}{2}|n{5}{2}|n{5}{2}|n{5}{2}|}
\hline
{Model} & {$MAE$} & {$RMSE$} & {$RMSE/\sigma$} & {$R^{2}$} \\
\hline
GB    & 0.4661487120273332 & 0.6690435174918778 & 0.7929034870155026     & 0.8255225713501024  \\
GKRR  & 0.7805128527272028 & 0.5160732879702644 & 0.5771993585227048     & 0.6001653484172635  \\
RF    & 0.6292607157871178 & 0.5280992856157458 & 0.3215649239826802     & 0.32895597666044296 \\
\hline
\end{tabular}
\label{p-vals}
\end{table}

\par
The optimal GB values of $MAE=2.18\pm 0.09$ $mm$, $RMSE=3.28\pm 0.13$ $mm$, $RMSE/\sigma = 0.70 \pm 0.04$, and $R^{2}=0.50 \pm 0.05$ are generally quite poor. In particular, the $RMSE/\sigma = 0.70$ suggests only modest improvement over simply guessing the mean of the dataset (which gives $RMSE/\sigma=1$) and $R^{2}=0.50$ is well below the qualitative guide of $R^{2} \approx 0.7$ that is often used to consider a result of significance. Therefore, these results suggest that it is unlikely that the CTs studied here can be used to provide a quantitative regression model for $D_{max}$ with a data set similar to that we have examined.

\par
\begin{figure}[H]
\centering
\includegraphics[width=0.75\textwidth]{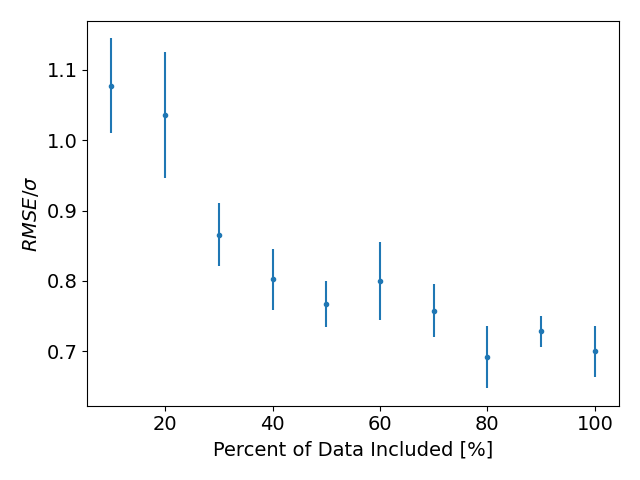}
\caption{$RMSE/\sigma$ decreases with an increase in the amount of data considered. The error bars are the standard error of the mean from all outer loop test sets in nested CV (see Sec.~\ref{Methods}).}
\label{learning_curve}
\end{figure}

One way to potentially improve the models for $D_{max}$ relative to those presented in this work is to add more data. To test the effect of the size of data on learning, a learning curve was produced using the best GB workflow for regression on $D_{max}$ (Figure~\ref{learning_curve}). After $50\%$ of the data are included, only minor improvements on $RMSE/\sigma$ were found, with a reduction from 0.77 at $50\%$ of the data to $0.70$ at $100 \%$ of the data. Consequently, a modest increase in the amount of data is unlikely to significantly improve learning $D_{max}$ using CTs. For example, assuming a linear extrapolation of rate of decrease from $50 \%$ to $100 \%$, the total database would have to grow by $300\%$ to get below a reasonable performance target of $RMSE/\sigma = 0.3$.

\par
\begin{figure}[H]
\centering
\includegraphics[width=0.75\textwidth]{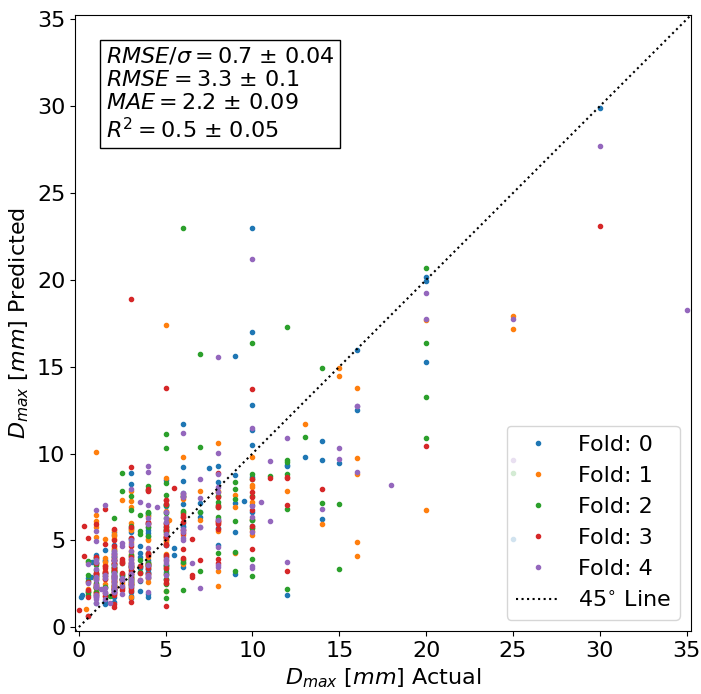}
\caption{The parity plot for test set prediction for the GB workflow is shown. The metrics in the annotation have SEM as the uncertainty.}
\label{parity}
\end{figure}

Another way to potentially improve the models is to explore a simpler classification in place of the full regression model. Here we consider classification into glasses with $D_{max}$ $<$ 4 $mm$ or $D_{max}$ $\ge$ 4 $mm$, where 4 $mm$ is the median of the dataset. The binary classification metrics for the GB classifier are shown in Table~\ref{binary}. Since the number of classes are near equal for the outer folds from nested CV, the baseline for F1 and receiver operator characteristic (ROC) area under the curve (AUC) scores is around 0.5. The scores for F1 above or equal to 4.0 $mm$, F1 below 4.0 $mm$, and ROC AUC are 0.77, 0.78, and 0.78 respectively, which represent a significant, although not outstanding, predictive ability. Although the regression metrics for predicting $D_{max}$ showed large uncertainties, we have shown that CTs are still potentially useful for classifying glasses above or below the median $D_{max}$. The success of classification can be understood by examining the parity plot for the GB regression workflow. As seen in Figure~\ref{parity}, predicted and actual $D_{max}$ values deviate significantly but show enough correlation that cases below 4.0 $mm$ tend to have predictions below 4.0 $mm$. Likewise, cases above or equal to 4.0 $mm$ tend to be predicted above or equal to 4.0 $mm$.

\par
We compared our workflow of GB regression to previous efforts by Xiong, et al. in Ref. \cite{Xiong2020} because they have the next largest database for metallic glass $D_{max}$. When applying 100-fold nested CV, our aggregate $RMSE=3.4$ $mm$ and $R^{2}=0.50$ which is comparable to their scores of $RMSE=2.89$ $mm$ and $R^{2}=0.60$ since no compositions with large residuals were excluded from our assessment. When an ordinary least squares model was fit and used to predict back $D_{max}$ from the terms in Equation~\ref{symbolic_function} for our dataset of 747 compositions, $R^{2}$ and $RMSE$ become $0.07$ and $4.6$ $mm$ respectively. Each term added a degree of freedom for fitting and we included the fitting intercept. Through further private communication with the authors, the general methodology to train the symbolic regression model may have suffered from data leakage.

\begin{table}[H]
\centering
\caption{The binary classification metrics for distinguishing metallic glasses above and below the median $D_{max}$ from nested CV are tabulated below.}
\npdecimalsign{.}
\nprounddigits{2}
\begin{tabular}{|c|n{5}{2}|n{5}{2}|n{5}{2}|}
\hline
{Metric}                             & {Mean} & {STDEV} & {SEM} \\
\hline
Cases for $D_{max} < 4$ $mm$    & 74.4          & 6.14817046     & 2.749545417  \\
Cases for $D_{max} \geq 4$ $mm$ & 75.0            & 6.557438524    & 2.93257566   \\
Accuracy                                    & 0.775131991   & 0.040543461    & 0.018131587  \\
F1 for $D_{max} < 4$ $mm$                 & 0.78171952    & 0.043599016    & 0.019498073  \\
F1 for $D_{max} \geq 4$ $mm$              & 0.766631603   & 0.042197508    & 0.018871299  \\
Precision for $D_{max} < 4$ $mm$          & 0.757241638   & 0.070499956    & 0.031528539  \\
Precision for $D_{max} \geq 4$ $mm$       & 0.800580869   & 0.05835977     & 0.026099283  \\
ROC AUC                                     & 0.777120335   & 0.040816002    & 0.018253471  \\
Recall for $D_{max} < 4$ $mm$             & 0.813050416   & 0.060951826    & 0.027258485  \\
Recall for $D_{max} \geq 4$ $mm$          & 0.741190254   & 0.075987653    & 0.033982712  \\
\hline
\end{tabular}
\label{binary}
\end{table}

\section{Conclusion}

\par
We have assessed the ability of features based on the characteristic temperatures (CTs) $T_{g}$, $T_{x}$, and $T_{l}$, to predict the critical casting diameter, $D_{max}$. We explored an extensive search of features based on powers and ratios of sums and differences of CTs, multiple machine learning models, and used nested cross validation to avoid data leakage when assessing the models. We found only weak ability for the models to predict  $D_{max}$ and found that to achieve significant improvement from increasing the database size would likely require a few multiples of the present database size. Given that we are already using the largest aggregated database to date, such an increase in amount of data would likely require a very large experimental effort or application of new high-throughput approaches.

\par
We also found that using just $T_{g}$, $T_{x}$, and $T_{l}$ directly was not statistically different than using features based on the powers and ratios of their sums and differences. These results suggest that further efforts adding terms within the examined space of features will not yield better predictive performance outside their training set compared to using the CTs directly. Some success was found in predicting $D_{max}$ above or below its median value from the CTs, suggesting that they can provide some valuable $D_{max}$ information. For example, models using these CTs could be used to screen small glassy samples and determine if larger glasses might be produced. Nevertheless, it appears that $D_{max}$ cannot be quantified with regression models built with the set of CTs examined. Previous linear models using CTs appear to have had more success when quantifying $R_{c}$ than $D_{max}$. This suggests that further exploration of $R_{c}$ models might be more fruitful than $D_{max}$ models. However, more complex models and more thorough assessment are limited by the limited amount of $R_{c}$ data, and more of such data would help in developing and assessing optimal CTs models.

\section*{Data Availability}

\par
The raw data required to reproduce these findings are available to download from [\url{https://petreldata.net/mdf/detail/voyles_mdf_dmref_glasses_v1.3/}]. The processed data required to reproduce these findings are available to download from [\url{https://petreldata.net/mdf/detail/schultz_gb_model_full_fit_v1.1/}].

\section*{Declaration of competing interest}

\par
The authors declare that they have no known competing financial interests or personal relationships that could have appeared to influence the work reported in this paper.

\section*{Acknowledgements}

\par
Lane E. Schultz is grateful for the financial support provided by the National Science Foundation (NSF), award number HRD-1612530 and from the University of Wisconsin– Madison Graduate Engineering Research Scholars (GERS) fellowship program. We gratefully acknowledge support from the NSF Designing Materials to Revolutionize and Engineer our Future (DMREF) program, Division of Materials Research (DMR), METAL \& METALLIC NANOSTRUCTUREs, award number \#1728933. Computational support was provided by the Extreme Science and Engineering Discovery Environment (XSEDE), which was supported by the National Science Foundation Grant No. OCI-1053575.

\section*{Appendix A}

\par
Scikit-learn metrics were used to assess the performance of ML models with the exception of $RMSE/\sigma$ \cite{scikit-learn}. Relevant metrics that have functional forms are defined in this section. Additional tables and a figure used in our work are presented here.

\setcounter{equation}{0}
\renewcommand{\theequation}{A\arabic{equation}}

\setcounter{table}{0}
\renewcommand{\thetable}{A\arabic{table}}

\setcounter{figure}{0}
\renewcommand{\thefigure}{A\arabic{figure}}

\begin{figure}[H]
\centering
\includegraphics[width=0.75\textwidth]{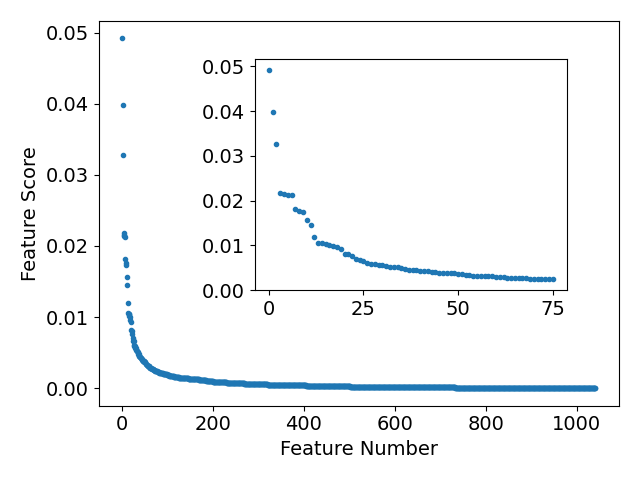}
\caption{The feature rankings for all PRSD functions of CTs for the GB model fit to all data.}
\label{features}
\end{figure}

\begin{table}[H]
\centering
\caption{The features scores for the full fit GB model. Higher score is better. Included are the top 1-50 features.}
\npdecimalsign{.}
\nprounddigits{2}
\begin{tabular}{|c|n{5}{2}|}
\hline
{Features} & {Scores} \\
\hline
{[}(tg-tx)\textasciicircum{}4{]}*1/{[}(tg-tl)\textasciicircum{}4{]} & 0.04917 \\
(tg-tx)*1/(tg-tl)                                                   & 0.03983 \\
{[}(tl-tx)\textasciicircum{}4{]}*1/{[}(tg-tx)\textasciicircum{}4{]} & 0.03274 \\
{[}tg\textasciicircum{}4{]}*1/{[}(tg-tl)\textasciicircum{}4{]}      & 0.02179 \\
(tl+tx)*1/{[}tl\textasciicircum{}2{]}                               & 0.02159 \\
{[}(tg-tl)\textasciicircum{}4{]}*1/{[}(tl-tx)\textasciicircum{}2{]} & 0.02137 \\
{[}(tg-tx)\textasciicircum{}2{]}*1/{[}(tl-tx)\textasciicircum{}2{]} & 0.02128 \\
(tg-tl)*1/(tg-tx)                                                   & 0.01810 \\
{[}(tl+tx)\textasciicircum{}4{]}*1/{[}tg\textasciicircum{}3{]}      & 0.01764 \\
{[}(tg-tl)\textasciicircum{}4{]}*1/{[}(tg-tx)\textasciicircum{}4{]} & 0.01737 \\
{[}(tg-tx)\textasciicircum{}4{]}*1/{[}(tl-tx)\textasciicircum{}4{]} & 0.01559 \\
{[}(tl+tx)\textasciicircum{}4{]}*1/{[}(tg-tx)\textasciicircum{}3{]} & 0.01452 \\
{[}tl\textasciicircum{}4{]}*{[}(tg-tx)\textasciicircum{}2{]}        & 0.01196 \\
{[}(tg-tl)\textasciicircum{}2{]}*1/{[}(tl-tx)\textasciicircum{}3{]} & 0.01059 \\
{[}(tl+tx)\textasciicircum{}3{]}*1/{[}tx\textasciicircum{}2{]}      & 0.01050 \\
{[}(tg-tx)\textasciicircum{}2{]}*1/(tl+tx)                          & 0.01041 \\
{[}tg\textasciicircum{}3{]}*1/{[}(tg+tx)\textasciicircum{}3{]}      & 0.01022 \\
{[}(tg+tx)\textasciicircum{}2{]}*1/{[}(tl-tx)\textasciicircum{}4{]} & 0.00997 \\
{[}(tg-tl)\textasciicircum{}2{]}*1/{[}(tg-tx)\textasciicircum{}2{]} & 0.00966 \\
{[}tl\textasciicircum{}3{]}*1/{[}(tg+tl)\textasciicircum{}3{]}      & 0.00925 \\
{[}(tg+tl)\textasciicircum{}4{]}*1/{[}(tg-tl)\textasciicircum{}4{]} & 0.00820 \\
{[}(tl+tx)\textasciicircum{}2{]}*1/{[}tx\textasciicircum{}4{]}      & 0.00809 \\
{[}(tg+tx)\textasciicircum{}3{]}*1/{[}(tg-tl)\textasciicircum{}4{]} & 0.00758 \\
tg*1/{[}(tg+tx)\textasciicircum{}3{]}                               & 0.00706 \\
{[}tx\textasciicircum{}3{]}*{[}(tl+tx)\textasciicircum{}4{]}        & 0.00670 \\
{[}(tl-tx)\textasciicircum{}4{]}*1/{[}(tg-tx)\textasciicircum{}2{]} & 0.00665 \\
(tl-tx)*1/{[}(tg-tx)\textasciicircum{}2{]}                          & 0.00608 \\
1/tl*1/{[}tg\textasciicircum{}4{]}                                  & 0.00596 \\
tl*1/(tg-tx)                                                        & 0.00590 \\
{[}(tg+tx)\textasciicircum{}2{]}*1/{[}(tg-tx)\textasciicircum{}2{]} & 0.00571 \\
{[}(tg-tx)\textasciicircum{}2{]}*1/{[}tx\textasciicircum{}3{]}      & 0.00560 \\
{[}tx\textasciicircum{}4{]}*1/{[}(tg-tx)\textasciicircum{}3{]}      & 0.00540 \\
{[}(tg-tx)\textasciicircum{}3{]}*1/{[}tl\textasciicircum{}2{]}      & 0.00524 \\
{[}(tg-tl)\textasciicircum{}3{]}*1/{[}tg\textasciicircum{}2{]}      & 0.00517 \\
1/(tg-tx)*1/{[}(tg+tl)\textasciicircum{}3{]}                        & 0.00514 \\
1/{[}tg\textasciicircum{}3{]}*1/{[}(tg-tl)\textasciicircum{}3{]}    & 0.00488 \\
{[}(tg+tx)\textasciicircum{}3{]}*1/{[}(tg+tl)\textasciicircum{}2{]} & 0.00485 \\
1/{[}tx\textasciicircum{}2{]}*1/{[}(tg-tx)\textasciicircum{}3{]}    & 0.00459 \\
{[}tl\textasciicircum{}2{]}*1/{[}(tl+tx)\textasciicircum{}4{]}      & 0.00449 \\
{[}(tg-tx)\textasciicircum{}3{]}*1/{[}tg\textasciicircum{}4{]}      & 0.00443 \\
{[}(tg-tx)\textasciicircum{}3{]}*1/tx                               & 0.00440 \\
(tg+tl)*1/{[}(tl-tx)\textasciicircum{}4{]}                          & 0.00431 \\
{[}tl\textasciicircum{}3{]}*1/{[}(tg+tx)\textasciicircum{}3{]}      & 0.00425 \\
{[}(tg-tl)\textasciicircum{}3{]}*{[}(tl-tx)\textasciicircum{}2{]}   & 0.00415 \\
1/{[}tx\textasciicircum{}4{]}*1/{[}(tg+tx)\textasciicircum{}4{]}    & 0.00401 \\
{[}(tl+tx)\textasciicircum{}4{]}*1/{[}(tg+tx)\textasciicircum{}2{]} & 0.00383 \\
tx*1/tg                                                             & 0.00383 \\
(tl-tx)*1/(tg-tl)                                                   & 0.00379 \\
(tl-tx)*{[}(tl+tx)\textasciicircum{}3{]}                            & 0.00377 \\
{[}(tg-tl)\textasciicircum{}4{]}*1/{[}(tl+tx)\textasciicircum{}3{]} & 0.00375 \\
\hline
\end{tabular}
\label{features_table_2}
\end{table}

\clearpage

\begin{equation}
R^{2} = 1-\frac{\sum_{i=1}^{n}(y_{i}-\hat{y_{i}})^{2}}{\sum_{i=1}^{n}(y_{i}-\bar{y})^{2}}
\end{equation}

\noindent
where:
\begin{itemize}
    \itemsep=0pt
    \item $R^{2}$ is the coefficient of determination
    \item $i$ is the sample number
    \item $n$ is the number of samples
    \item $y_{i}$ is the true target value for a case $i$
    \item $\hat{y_{i}}$  is the predicted target value for a case $i$
    \item $\bar{y}$ is the mean of true target values
\end{itemize}

\begin{equation}
MAE = \frac{1}{n} \sum_{i=0}^{n-1} |y_{i}-\hat{y_{i}}|
\end{equation}

\noindent
where:
\begin{itemize}
    \itemsep=0pt
    \item $MAE$ is the mean absolute error
    \item $i$ is the sample number
    \item $n$ is the number of samples
    \item $y_{i}$ is the true target value for a case $i$
    \item $\hat{y_{i}}$  is the predicted target value for a case $i$
\end{itemize}

\begin{equation}
MSE = \frac{1}{n} \sum_{i=0}^{n-1} (y_{i}-\hat{y_{i}})^{2}
\end{equation}

\noindent
where:
\begin{itemize}
    \itemsep=0pt
    \item $MSE$ is the mean squared error
    \item $i$ is the sample number
    \item $n$ is the number of samples
    \item $y_{i}$ is the true target value for a case $i$
    \item $\hat{y_{i}}$  is the predicted target value for a case $i$
\end{itemize}

\begin{equation}
RMSE = \sqrt{MSE}
\end{equation}

\noindent
where:
\begin{itemize}
    \itemsep=0pt
    \item $RMSE$ is the root mean squared error
    \item $MSE$ is the mean squared error
\end{itemize}

\pagebreak

\begin{equation}
RMSE/\sigma = \frac{RMSE}{\sigma}
\end{equation}

\noindent
where:
\begin{itemize}
    \itemsep=0pt
    \item $RMSE/\sigma$ is the ratio between RMSE and $\sigma$
    \item $RMSE$ is the root mean squared error
    \item $\sigma$ is the standard deviation in the true target values
\end{itemize}

\begin{equation}
accuracy = \frac{1}{n} \sum_{i=0}^{n-1} 1(y_{i}=\hat{y_{i}})
\end{equation}

\noindent
where:
\begin{itemize}
    \itemsep=0pt
    \item $accuracy$ is the accuracy
    \item $i$ is the sample number
    \item $n$ is the number of samples
    \item $y_{i}$ is the true target value for a case $i$
    \item $\hat{y_{i}}$  is the predicted target value for a case $i$
\end{itemize}

\begin{equation}\label{precision}
precision = \frac{tp}{tp+fp}
\end{equation}

\noindent
where:
\begin{itemize}
    \itemsep=0pt
    \item $precision$ is the precision
    \item $tp$ is the number of true positives
    \item $fp$ is the number of false positives
\end{itemize}

\begin{equation}\label{recall}
recall = \frac{tp}{tp+fn}
\end{equation}

\noindent
where:
\begin{itemize}
    \itemsep=0pt
    \item $recall$ is the recall
    \item $tp$ is the number of true positives
    \item $fn$ is the number of false negatives
\end{itemize}

\begin{equation}
F_{1} = 2 \cdot \frac{precision \cdot recall}{precision+recall}
\end{equation}

\noindent
where:
\begin{itemize}
    \itemsep=0pt
    \item $F_{1}$ is the harmonic mean between precision and recall
    \item $precision$ is defined in Equation~\ref{precision}
    \item $recall$ is defined in Equation~\ref{recall}
\end{itemize}

\begin{table}[H]
\centering
\caption{The mean and standard deviation for the outer loops in nested cross validation for the generated set of features for GB models.}
\npdecimalsign{.}
\nprounddigits{2}
\begin{tabular}{|c|c|c|n{5}{2}|n{5}{2}|n{5}{2}|}
\hline
{Metric} & {$Log_{10}$} & {PCA} & {Mean} & {STDEV} & {SEM} \\
\hline
$MAE$         & False & False & 2.177763262         & 0.20983286286683156  & 0.09384010905672535  \\
$MAE$         & False & True  & 2.891776064516585   & 0.31894504947493196  & 0.14263656234259628  \\
$MAE$         & True  & False & 4.6099479064288875  & 0.16371151120692565  & 0.07321401355158087  \\
$MAE$         & True  & True  & 4.628987817983644   & 0.31625533409209056  & 0.14143368505536424  \\
$RMSE$        & False & False & 3.2773837047999996  & 0.2910760810024851   & 0.13017318074915837  \\
$RMSE$        & False & True  & 4.0562356953104155  & 0.48086753706526913  & 0.2150505002101683   \\
$RMSE$        & True  & False & 6.5062684574828085  & 0.4074756753961659   & 0.1822286618726931   \\
$RMSE$        & True  & True  & 6.53377137326657    & 0.5443577776409249   & 0.2434441989771646   \\
$RMSE/\sigma$ & False & False & 0.6999509862        & 0.0801319762526573   & 0.035836109214468115 \\
$RMSE/\sigma$ & False & True  & 0.857365730626731   & 0.10079693444412109  & 0.04507775946812895  \\
$RMSE/\sigma$ & True  & False & 1.3682035770070058  & 0.037107466404276084 & 0.016594963470550204 \\
$RMSE/\sigma$ & True  & True  & 1.3800803923241676  & 0.061462271163876564 & 0.02748676327479062  \\
$R^{2}$       & False & False & 0.5049317097999999  & 0.11294360877779022  & 0.050509917370256174 \\
$R^{2}$       & False & True  & 0.2567959863522256  & 0.17587424784263436  & 0.07865335473355523  \\
$R^{2}$       & True  & False & -0.8730825993851212 & 0.1014556412365782   & 0.045372342101163934 \\
$R^{2}$       & True  & True  & -0.9076439778989254 & 0.172174675912279    & 0.07699885586877028  \\
\hline
\end{tabular}
\label{gen_mean_metrics_gb}
\end{table}

\begin{table}[H]
\centering
\caption{The mean and standard deviation for the outer loops in nested cross validation for the generated set of features for GKRR models.}
\npdecimalsign{.}
\nprounddigits{2}
\begin{tabular}{|c|c|c|n{5}{2}|n{5}{2}|n{5}{2}|}
\hline
{Metric} & {$Log_{10}$} & {PCA} & {Mean} & {STDEV} & {SEM} \\
\hline
$MAE$         & False & False & 2.3613433362790883  & 0.13924765030617506  & 0.062273442358345366 \\
$MAE$         & False & True  & 2.3985362824258245  & 0.2357724851494426   & 0.10544066080364266  \\
$MAE$         & True  & False & 4.6271588021869805  & 0.2320277928638923   & 0.10376598350258076  \\
$MAE$         & True  & True  & 4.626462599674582   & 0.43853454804359093  & 0.19611861198152333  \\
$RMSE$        & False & False & 3.5699478301399226  & 0.24978766626721957  & 0.11170844034290682  \\
$RMSE$        & False & True  & 3.584032835204055   & 0.36655857442278783  & 0.16392997802895387  \\
$RMSE$        & True  & False & 6.522015528418898   & 0.578414544745079    & 0.2586748482449181   \\
$RMSE$        & True  & True  & 6.518601651691111   & 0.6137571786467962   & 0.27448055462654375  \\
$RMSE/\sigma$ & False & False & 0.7553606795917878  & 0.06282084870501099  & 0.02809433762172684  \\
$RMSE/\sigma$ & False & True  & 0.7522565114170061  & 0.0865628482612408   & 0.038712082607626784 \\
$RMSE/\sigma$ & True  & False & 1.3854035599885715  & 0.0886005402698009   & 0.03962336617729647  \\
$RMSE/\sigma$ & True  & True  & 1.3772221668571007  & 0.05334841742827817  & 0.0238581375723329   \\
$R^{2}$       & False & False & 0.4262730765010182  & 0.09359081621733333  & 0.04185508542632941  \\
$R^{2}$       & False & True  & 0.4281156396714369  & 0.1331360088431875   & 0.05954023320527607  \\
$R^{2}$       & True  & False & -0.9256230686178875 & 0.2484332905553236   & 0.111102745111132    \\
$R^{2}$       & True  & True  & -0.8990177397962494 & 0.14407224861359746  & 0.06443106831425074  \\
\hline
\end{tabular}
\label{gen_mean_metrics_gkrr}
\end{table}

\begin{table}[H]
\centering
\caption{The mean and standard deviation for the outer loops in nested cross validation for the generated set of features for LASSO models.}
\npdecimalsign{.}
\nprounddigits{2}
\begin{tabular}{|c|c|c|n{5}{2}|n{5}{2}|n{5}{2}|}
\hline
{Metric} & {$Log_{10}$} & {PCA} & {Mean} & {STDEV} & {SEM} \\
\hline
$MAE$         & False & False & 2.7897354139571     & 0.2320999806498647   & 0.13400298630710633  \\
$MAE$         & False & True  & 2.750735093038043   & 0.116055298813525    & 0.06700455801087112  \\
$MAE$         & True  & False & 4.763390560866818   & 0.18163581320280855  & 0.1048674856471181   \\
$MAE$         & True  & True  & 4.570178431425023   & 0.404904529582222    & 0.23377173915039468  \\
$RMSE$        & False & False & 3.9572936528685325  & 0.4957877757482487   & 0.2862432057891772   \\
$RMSE$        & False & True  & 3.9688634753918017  & 0.223921701441046    & 0.12928125460438694  \\
$RMSE$        & True  & False & 6.814922722406348   & 0.3238372486180373   & 0.18696752266325162  \\
$RMSE$        & True  & True  & 6.664218202395705   & 0.67697502016161     & 0.3908517101249579   \\
$RMSE/\sigma$ & False & False & 0.8648449751236899 & 0.028263517712638183 & 0.01631794955963741 \\
$RMSE/\sigma$ & False & True  & 0.8148350498666646  & 0.04149065128203427  & 0.023954638686535377 \\
$RMSE/\sigma$ & True  & False & 1.3995895937524139  & 0.04603859480975342  & 0.026580395106523245 \\
$RMSE/\sigma$ & True  & True  & 1.3602218043301997  & 0.03180678290260179  & 0.01836365467087313  \\
$R^{2}$       & False & False & 0.2515106180476426  & 0.04876304786554907  & 0.02815335881168136  \\
$R^{2}$       & False & True  & 0.3348961920795854  & 0.06837794396503712  & 0.039478024354847326 \\
$R^{2}$       & True  & False & -0.9602640657480853 & 0.12791422430041     & 0.07385131183302389  \\
$R^{2}$       & True  & True  & -0.850877804601046  & 0.08652398446786332  & 0.049954645723879885 \\
\hline
\end{tabular}
\label{gen_mean_metrics_lasso}
\end{table}

\begin{table}[H]
\centering
\caption{The mean and standard deviation for the outer loops in nested cross validation for the generated set of features for RF models.}
\npdecimalsign{.}
\nprounddigits{2}
\begin{tabular}{|c|c|c|n{5}{2}|n{5}{2}|n{5}{2}|}
\hline
{Metric} & {$Log_{10}$} & {PCA} & {Mean} & {STDEV} & {SEM} \\
\hline
$MAE$         & False & False & 2.2435760886332163  & 0.1904022433668076   & 0.08515047184732805  \\
$MAE$         & False & True  & 3.1983609342424866  & 0.32854955881419057  & 0.14693182949721906  \\
$MAE$         & True  & False & 4.611418298883483   & 0.17341416036785093  & 0.07755317016871292  \\
$MAE$         & True  & True  & 4.641684764332056   & 0.30317244892695566  & 0.1355828409411512   \\
$RMSE$        & False & False & 3.3024829928718544  & 0.3271818757146302   & 0.14632018302076014  \\
$RMSE$        & False & True  & 4.3019345290653686  & 0.4777789853641467   & 0.21366925789902183  \\
$RMSE$        & True  & False & 6.512707784779812   & 0.4216083217081163   & 0.18854897344378965  \\
$RMSE$        & True  & True  & 6.519536857631695   & 0.66234486324453     & 0.2962096277525142   \\
$RMSE/\sigma$ & False & False & 0.6979387534177431  & 0.04267062132404659  & 0.01908288198454405  \\
$RMSE/\sigma$ & False & True  & 0.9203845318995703  & 0.1434376447785077   & 0.06414726485144219  \\
$RMSE/\sigma$ & True  & False & 1.372737087333297   & 0.059896939007054326 & 0.026786725452786445 \\
$RMSE/\sigma$ & True  & True  & 1.3817012508845425  & 0.05315296734904437  & 0.023770729639658    \\
$R^{2}$       & False & False & 0.5114248709383427  & 0.06096675635827984  & 0.02726516231695625  \\
$R^{2}$       & False & True  & 0.13643282708832485 & 0.26168252145192     & 0.11702798129800801  \\
$R^{2}$       & True  & False & -0.8872772255822354 & 0.16524626632160866  & 0.07390037690463021  \\
$R^{2}$       & True  & True  & -0.9113585370463164 & 0.1447115057281901   & 0.06471695278691665  \\
\hline
\end{tabular}
\label{gen_mean_metrics_rf}
\end{table}

\begin{table}[H]
\centering
\caption{The mean and standard deviation for the outer loops in nested cross validation for the three characteristic temperatures feature set for GB models.}
\npdecimalsign{.}
\nprounddigits{2}
\begin{tabular}{|c|c|c|n{5}{2}|n{5}{2}|n{5}{2}|}
\hline
{Metric} & {$Log_{10}$} & {PCA} & {Mean} & {STDEV} & {SEM} \\
\hline
$MAE$         & False & False & 2.3038071245688037    & 0.30011688498538514   & 0.13421635120456143   \\
$MAE$         & False & True  & 2.524718596674614     & 0.19230111768077293   & 0.085999674256679     \\
$MAE$         & True  & False & 4.601822106081367     & 0.27307666878069287   & 0.12212359889256477   \\
$MAE$         & True  & True  & 4.598192398047869     & 0.59994270300391      & 0.26830253330434195   \\
$RMSE$        & False & False & 3.3802678099674837    & 0.4260500469805222    & 0.19053537337308532   \\
$RMSE$        & False & True  & 3.741999268010671     & 0.30571238918017324   & 0.1367187367541477    \\
$RMSE$        & True  & False & 6.5125443479459       & 0.4571914843794533    & 0.20446224756129816   \\
$RMSE$        & True  & True  & 6.49088300614333      & 0.8136718301282689    & 0.3638851047086941    \\
$RMSE/\sigma$ & False & False & 0.7118315420707944    & 0.05546396195757342   & 0.024804237847719294  \\
$RMSE/\sigma$ & False & True  & 0.80396121808146      & 0.10459226406489568   & 0.04677508247394304   \\
$RMSE/\sigma$ & True  & False & 1.3742499468614846    & 0.06201868166817488   & 0.027735597616991816  \\
$RMSE/\sigma$ & True  & True  & 1.3773898889084426    & 0.04124391674851844   & 0.018444840301605866  \\
$R^{2}$       & False & False & 0.49083485485229      & 0.07903278892495835   & 0.03534453769751988   \\
$R^{2}$       & False & True  & 0.34489472645919805   & 0.162449815409788     & 0.07264976603771577   \\
$R^{2}$       & True  & False & -0.8916399699494795   & 0.16990642267811673   & 0.07598446218441617   \\
$R^{2}$       & True  & True  & -0.8985637546022194   & 0.11412764667312905   & 0.051039435214638856  \\
\hline
\end{tabular}
\label{raw_mean_metrics_gb}
\end{table}

\begin{table}[H]
\centering
\caption{The mean and standard deviation for the outer loops in nested cross validation for the three characteristic temperatures feature set for GKRR models.}
\npdecimalsign{.}
\nprounddigits{2}
\begin{tabular}{|c|c|c|n{5}{2}|n{5}{2}|n{5}{2}|}
\hline
{Metric} & {$Log_{10}$} & {PCA} & {Mean} & {STDEV} & {SEM} \\
\hline
$MAE$         & False & False & 2.334231470029514     & 0.15751591248921323   & 0.07044325757275778   \\
$MAE$         & False & True  & 2.293049596820203     & 0.3714865640347572    & 0.16613384198190914   \\
$MAE$         & True  & False & 4.611874076543681     & 0.39412608761787093   & 0.1762585447239195    \\
$MAE$         & True  & True  & 4.623022808177451     & 0.23594085449030877   & 0.10551595786194337   \\
$RMSE$        & False & False & 3.44083555102022      & 0.34158754807254466   & 0.15276259555153743   \\
$RMSE$        & False & True  & 3.5114785201003142    & 0.7658328627410851    & 0.3424908680984664    \\
$RMSE$        & True  & False & 6.486768004997583     & 0.8026898664737449    & 0.3589738212571046    \\
$RMSE$        & True  & True  & 6.512991190794213     & 0.4994928202270974    & 0.22337998006017434   \\
$RMSE/\sigma$ & False & False & 0.7290478570924416    & 0.0791376522556322    & 0.03539143400466663   \\
$RMSE/\sigma$ & False & True  & 0.7464754873159645    & 0.05781182829876647   & 0.02585423559591757   \\
$RMSE/\sigma$ & True  & False & 1.3799482513569807    & 0.04664955420501237   & 0.020862314864493765  \\
$RMSE/\sigma$ & True  & True  & 1.3753792110486085    & 0.054960896068533355  & 0.024579259942708303  \\
$R^{2}$       & False & False & 0.46347900766529226   & 0.11989224667656069   & 0.05361744270879259   \\
$R^{2}$       & False & True  & 0.44010058084339654   & 0.08591243207322943   & 0.03842120764561484   \\
$R^{2}$       & True  & False & -0.9059981211492094   & 0.1289076770643069    & 0.05764926574747614   \\
$R^{2}$       & True  & True  & -0.8940845342620175   & 0.1500738595551005    & 0.0671150703221922    \\
\hline
\end{tabular}
\label{raw_mean_metrics_gkrr}
\end{table}

\begin{table}[H]
\centering
\caption{The mean and standard deviation for the outer loops in nested cross validation for the three characteristic temperatures feature set for LASSO models.}
\npdecimalsign{.}
\nprounddigits{2}
\begin{tabular}{|c|c|c|n{5}{2}|n{5}{2}|n{5}{2}|}
\hline
{Metric} & {$Log_{10}$} & {PCA} & {Mean} & {STDEV} & {SEM} \\
\hline
$MAE$         & False & False & 3.37233852724525      & 0.23249185774362155   & 0.10397351962598973   \\
$MAE$         & False & True  & 3.3682781346295116    & 0.23352320777173494   & 0.1044347533802813    \\
$MAE$         & True  & False & 4.617597524915579     & 0.5863801362459856    & 0.2622371690603224    \\
$MAE$         & True  & True  & 4.617902891807086     & 0.36526141866880507   & 0.16334987234029177   \\
$RMSE$        & False & False & 4.762245844421407     & 0.5964553268894798    & 0.266742931293347     \\
$RMSE$        & False & True  & 4.75368807746761      & 0.5936768668339425    & 0.26550036618195716   \\
$RMSE$        & True  & False & 6.588507348768478     & 0.950506706039091     & 0.42507952155456347   \\
$RMSE$        & True  & True  & 6.612436994276349     & 0.719769486582116     & 0.3218907000255468    \\
$RMSE/\sigma$ & False & False & 1.0058979329788433    & 0.007024782716561641  & 0.0031415783362794933 \\
$RMSE/\sigma$ & False & True  & 1.0014018312136275    & 0.0013585664579173744 & 0.0006075693903708713 \\
$RMSE/\sigma$ & True  & False & 1.4053273882450334    & 0.04838386823737272   & 0.021637923678631668  \\
$RMSE/\sigma$ & True  & True  & 1.404645219407233     & 0.07011606243184518   & 0.031356856382445     \\
$R^{2}$       & False & False & -0.011870129628881499 & 0.014188739465995704  & 0.006345397192200092  \\
$R^{2}$       & False & True  & -0.002807104120262904 & 0.0027218314273378097 & 0.0012172400189645243 \\
$R^{2}$       & True  & False & -0.9768178671160959   & 0.13646670788638526   & 0.06102976709991282   \\
$R^{2}$       & True  & True  & -0.9769612021723516   & 0.19765217385272746   & 0.08839273932706102   \\
\hline
\end{tabular}
\label{raw_mean_metrics_lasso}
\end{table}

\begin{table}[H]
\centering
\caption{The mean and standard deviation for the outer loops in nested cross validation for the three characteristic temperatures feature set for RF models.}
\npdecimalsign{.}
\nprounddigits{2}
\begin{tabular}{|c|c|c|n{5}{2}|n{5}{2}|n{5}{2}|}
\hline
{Metric} & {$Log_{10}$} & {PCA} & {Mean} & {STDEV} & {SEM} \\
\hline
$MAE$         & False & False & 2.305614888888889     & 0.20031901870439436   & 0.08958538860181553   \\
$MAE$         & False & True  & 2.4003764990476193    & 0.23936035478121725   & 0.1070452048818537    \\
$MAE$         & True  & False & 4.592010228922547     & 0.41586720214205014   & 0.18598146672045404   \\
$MAE$         & True  & True  & 4.597990493722615     & 0.3805699094471921    & 0.1701960375429722    \\
$RMSE$        & False & False & 3.4302331276254874    & 0.28323114274599787   & 0.12666481770499952   \\
$RMSE$        & False & True  & 3.4983717029069035    & 0.5477358720234506    & 0.24495492871191218   \\
$RMSE$        & True  & False & 6.481530356763808     & 0.6265544933749427    & 0.28020368775886273   \\
$RMSE$        & True  & True  & 6.4832565382244995    & 0.7841601638371697    & 0.3506870863174567    \\
$RMSE/\sigma$ & False & False & 0.7316363000408768    & 0.05669315669454219   & 0.02535395044560872   \\
$RMSE/\sigma$ & False & True  & 0.7384329254627229    & 0.06864987473931923   & 0.03070115731279269   \\
$RMSE/\sigma$ & True  & False & 1.3674189057495327    & 0.03280359421753608   & 0.014670213315345938  \\
$RMSE/\sigma$ & True  & True  & 1.382700302448057     & 0.07243350579540234   & 0.03239324856142892   \\
$R^{2}$       & False & False & 0.4621370332497028    & 0.08582916981119945   & 0.03838397163004295   \\
$R^{2}$       & False & True  & 0.450946570351185     & 0.10369773227288903   & 0.046375035694950724  \\
$R^{2}$       & True  & False & -0.8706953244361205   & 0.08997172820956152   & 0.040236580065943     \\
$R^{2}$       & True  & True  & -0.9160574165993982   & 0.19928410943005304   & 0.08912256310422109   \\
\hline
\end{tabular}
\label{raw_mean_metrics_rf}
\end{table}

\clearpage
\printbibliography[heading=bibintoc]

\end{document}